\documentclass{JHEP}
\usepackage{epsfig}
\title{Fixing the conformal window in QCD\footnote{Extended version of a talk given at the QCDNET2000 network meeting, Paris, 
September 11-14 2000 (http://qcd.th.u-psud.fr/QCDNET/).}}
\author{G.\ Grunberg\thanks{Research
supported in part by the EC program ``Training and 
Mobility of 
Researchers'', Network ``QCD and Particle Structure'', contract 
ERBFMRXCT980194.}\\
        Centre de Physique Th\'eorique de l' Ecole  
Polytechnique (CNRS UMR C7644),\\
        91128 Palaiseau Cedex, France\\
        E-mail: \email{georges.grunberg@cpht.polytechnique.fr}}
\abstract{A physical characterization of Landau singularities is emphasized, which should trace the lower boundary
$N_f^*$ of the conformal window in QCD and supersymmetric QCD. A natural way to disentangle ``perturbative'' from ``non-perturbative'' contributions 
 below $N_f^*$ is suggested. Assuming an infrared fixed point persists in the perturbative part of the
QCD coupling in some range below $N_f^*$ leads to the condition $\gamma(N_f^*)=1$, where $\gamma$ is the critical
 exponent. Using the Banks-Zaks expansion, one gets 
 $4\leq N_f^*\leq 6$. This result is incompatible with the existence of an analogue of Seiberg duality
 in QCD. The presence of a negative ultraviolet fixed point is required both in QCD and in supersymmetric QCD
 to preserve causality 
within the conformal window. Evidence for the existence of such a fixed point in QCD is provided.}
\preprint{CPTh/PC 087.0900}
\begin{document}

\section{Introduction}
The notion of an infrared (IR) finite  coupling has been used  extensively in recent years, especially
in connection \cite{Dok} with the phenomenology of power corrections in QCD. The present investigation  is motivated by the
desire to understand better the theoretical background behind such an assumption. In particular, given
an IR finite coupling $\alpha$, does it remain  finite within perturbation
theory itself (such as the two-loop coupling  with opposite signs one and two loop beta function coefficients), or does one need a
non-perturbative contribution $\delta\alpha$  to cancell ($\alpha=\alpha_{PT}+\delta\alpha$) the Landau singularities present in its
  perturbative part $\alpha_{PT}$? The answer I shall suggest is a mixed one: the perturbative
part of the
QCD coupling is {\em always} IR finite but, below the so called ``conformal window'' (the range of $N_f$ values where the theory
is scale invariant at large distances and flows to a non-trivial IR fixed point), 
one still needs a $\delta\alpha$
term since the perturbative coupling is no more causal there, despite being IR finite. As the main outcome, one obtains
 an equation to determine
the lower boundary of the conformal window in QCD. The plan of the paper  is as follows. In section 2 I review the evidence
and present a formal argument
for the existence of Landau singularities in the perturbative coupling. A more physical argument, relating Landau
singularities to the very existence of the conformal window and a two-phase structure of QCD is given in section 3, which
also suggests a clean way to disentangle ``perturbative'' from ``non-perturbative'' below the conformal window. In section
4, two scenarios for causality breaking are described. In section 5, an equation to determine the bottom of the conformal
window in QCD is suggested, and is solved through the Banks-Zaks expansion in section 6. Section 7 gives evidence, through
a modified Banks-Zaks expansion, for the existence of a {\em negative} ultraviolet (UV) fixed point in QCD, necessary for the
consistency of the present approach.
 Section 8 contains the conclusions.

\section{Evidence for Landau singularities in the perturbative coupling}
The only present evidence for a Landau singularity in the  perturbative 
{\em renormalized}\footnote{In QED, the well established ``triviality'' property gives only direct
 evidence \cite{G-qed}
for a singularity in the {\em bare} coupling constant.}
 coupling is still
the old Landau-Pomeranchuk leading log QED calculation, now reformulated in QCD as a $N_f\rightarrow -\infty$
(``large $\beta_0$'') limit.  In this limit, the perturbative coupling is one-loop:
$\alpha_{PT}(k^2)=1/ \beta_0 \log(k^2/ \Lambda^2)$,
where $\Lambda$ is the Landau  pole. The question is whether there is a singularity
at {\em finite} $N_f$. Some light on this problem can be shed by considering further the $N_f$ dependence. Indeed,
 another (conflicting) piece of information is available at the other end of the spectrum,
 around the value $N_f=N_f^0=16.5$ (I consider $N_c=3$) where the one loop
 coefficient $\beta_0={1\over 4}(11-{2\over 3}N_f)$
of the beta function vanishes (``small $\beta_0$'' limit). For $N_f$ slightly below $16.5$ 
 a weak coupling (Banks-Zaks) 
IR fixed point develops \cite{BZ,White,G-bz}  and the perturbative coupling is causal,
 i.e. there are no Landau singularities in 
the whole first sheet of the complex $k^2$ plane. Can then the perturbative coupling remain causal down to 
$N_f=-\infty$? I shall assume that a Landau singularity cannot arise ``spontaneously''
in the limit, i.e. that
``the limit of a sequence of causal couplings must itself be causal''. In such a case, the existence of 
a Landau pole
at $N_f\rightarrow -\infty$ implies the existence of a {\em finite} value $N_f^*$  below which
Landau singularities appear on the first sheet of the complex $k^2$ plane and perturbative causality
is lost, which is the common wisdom (at $N_f^*$ itself, according to the above philosophy, the 
coupling must still be causal). The range $N_f^*<N_f<N_f^0$ where the {\em perturbative} coupling is causal
and flows to a finite IR fixed point
is taken as the definition of  the ``conformal window'' for the sake of the present discussion. Let us now refine this
definition, and give
 a more physical argument for the existence of Landau singularities, which  illuminates
their physical meaning.

\section{Landau singularities and conformal window}
I assume the existence of a two-phase structure in QCD as the number of flavors $N_f$ is varied. 

i) For $N_f^*<N_f<N_f^0$  (the conformal window) the theory is scale invariant at large distances, and the vacuum is 
``perturbative'', in the sense there is no confinement nor chiral symmetry breaking. Conformal window amplitudes (generically
noted as $D_{\overline{PT}}(Q^2)$, where $Q$ stands for an external scale) are in this generalized sense ``perturbative'',
 i.e. could in principle be determined from information
contained in perturbation theory to all orders (although they should also  include contributions from all instanton sectors):
 this motivates the subscript $\overline{PT}$.

ii) For $0<N_f<N_f^*$ there is a phase transition to a non-trivial vacuum, with confinement and chiral symmetry breaking, as expected in 
standard QCD.

A direct, {\em physical} motivation for Landau singularities can now be given: they trace the lower boundary $N_f=N_f^*$
of the conformal window. This statement is implied from the following two postulates:

1) Conformal window amplitudes $D_{\overline{PT}}(Q^2)$ can be analytically continued in $N_f$ below the bottom  $N_f^*$
of the conformal window.

2) For $N_f<N_f^*$, the (analytically continued) conformal window amplitudes $D_{\overline{PT}}(Q^2)$ must
 {\em differ} from the
 {\em full}  QCD amplitude $D(Q^2)$, since one enters
a new phase, i.e. we have

\begin{equation}D(Q^2)=D_{\overline{PT}}(Q^2)+D_{\overline{NP}}(Q^2)\label{eq:PT-NP}
\end{equation}
(whereas $D(Q^2)\equiv D_{\overline{PT}}(Q^2)$ within the conformal window).
 Assuming QCD to be a {\em unique} theory at given $N_f$, $D_{\overline{PT}}(Q^2)$ cannot provide a consistent
solution if $N_f<N_f^*$: this must be signalled by the appearance of unphysical Landau singularities 
in $D_{\overline{PT}}(Q^2)$. $N_f^*$ should thus coincide with the value of $N_f$ below which (first sheet) Landau 
singularities  appear in
$D_{\overline{PT}}(Q^2)$. The occurence of a ``genuine'' non-perturbative component $D_{\overline{NP}}(Q^2)$  is then 
necessary below $N_f^*$ in order to cancell the Landau singularities present in $D_{\overline{PT}}(Q^2)$.
 If these assumptions are correct, they provide an interesting connection between information contained
in principle in ``perturbation theory'' (over all instanton sectors), which fix the structure of the conformal window
 amplitudes and ``genuine'' non-perturbative
phenomena, which fix the bottom of the conformal window. In addition, eq.(\ref{eq:PT-NP}) provide a neat way to disentangle
 the ``perturbative'' from the genuine ``non-perturbative'' part of an amplitude, for instance the part of the gluon
 condensate related to renormalons from the one reflecting
the presence of the non-trivial vacuum. Note also $D_{\overline{PT}}(Q^2)$ and $D_{\overline{NP}}(Q^2)$ are separately
free of renormalons ambiguities, but contain Landau singularities below $N_f^*$, so the renormalon and Landau
singularity problems are also disentangled. In order to get a precise condition to determine $N_f^*$, one needs  to look
in more details how causality can be broken in the perturbative coupling.

\section{Scenarios for causality breaking}
There are two main  scenarios:

i) The ``standard'' one where the IR fixed point present within the conformal window just disappears when
$N_f<N_f^*$ while a real, space-like Landau singularity is generated in the perturbative coupling. For instance, two
 simple zeroes of the beta function can merge into a  double zero when $N_f=N_f^*$  before moving to the complex plane 
(a plausible scenario \cite{Gar-Gru-conformal} in
supersymmetric QCD (SQCD)). 

ii) Alternatively, it is possible for the fixed point {\em to be still present}\footnote{This assumption is consistent
with the suggestion  \cite{Stev} that the {\em perturbative} coupling has a non-trivial IR fixed point down
 to $N_f=2$ in QCD.
 However
the full non-perturbative coupling must still differ by a $\delta\alpha$ term, since the perturbative coupling is
non-causal below $N_f=N_f^*$.}
 in the {\em perturbative part} of the coupling  in some range of $N_f$ {\em below}
 $N_f^*$. The motivation behind this assumption is the observation \cite{Stev,Gar-Kar,Gar-Gru-conformal} that for QCD effective
charges associated to Euclidean correlators (the only ones for which the notion of $k^2$ plane analyticity makes sense) 
the Banks-Zaks expansion in QCD (as opposed to SQCD \cite{Gar-Gru-conformal}) seems to converge down to fairly small values of $N_f$.
In this case
there can be no space-like Landau singularity, and 
causality must be violated by the appearance of
{\em complex} Landau singularities  on the first sheet of the  $k^2$ plane. The  example below suggests 
that they arise 
as the result of the continuous migration  to the first sheet, through
 the time-like cut, of some second sheet  singularities already
present when $N_f>N_f^*$. I shall assume that this is the scenario which prevails in  QCD. As the simplest example, 
consider the two-loop coupling, which satisfies the renormalization group (RG) equation
$d\alpha/ d\log k^2=-\beta_0 \alpha^2-\beta_1 \alpha^3$.
If $\beta_0 >0$ but $\beta_1 <0$, there is an IR fixed point at $\alpha_{IR}=-\beta_0/ \beta_1$.
 It has been shown  \cite{Uraltsev,Gar-Gru-Kar,Gar-Gru-conformal} that this coupling has a pair of complex conjugate Landau singularities 
  on the second (or higher)
sheet if
 \begin{equation}0<\gamma_{2-loop}=-{\beta_0^2\over \beta_1}<1\label{eq:gamma-2loop}
\end{equation}
where $\gamma_{2-loop}$ is the 2-loop critical exponent (see
below).
For $\gamma_{2-loop}>1$, the second sheet singularities move to the first sheet through the time-like cut,
 which is reached when $\gamma_{2-loop}=1$. The latter condition thus determines the bottom of the 
conformal window in this model. Note that in the limit $\beta_1\rightarrow 0^-$ where $\gamma_{2-loop}=+\infty$,
 one gets the one loop coupling and the complex conjugate singularities collapse to  a  space-like Landau pole. This 
limit is thus the analogue of the $N_f\rightarrow -\infty$ limit in  QCD. 

\section{An equation to determine the bottom of the conformal window in QCD}
Assuming from now on that the second scenario described above applies, i.e. that the IR fixed point persists in some range of 
$N_f$ {\em below} $N_f^*$, one needs some information on the location of Landau singularities in coupling constant space to derive an 
equation for $N_f^*$.  I shall assume throughout that there are no {\em complex} (in $\alpha$ space)
Landau singularities (such as complex poles in the beta function), i.e. that the Landau singularities originate only
from the $\alpha<0$ or from the $\alpha>\alpha_{IR}$ regions, and argue that the condition
\begin{equation}0<\gamma<1\label{eq:gamma-causal}\end{equation}
is both necessary and sufficient for causality in QCD. Consequently, the
lower boundary of the conformal window is obtained from the equation
\begin{equation}\gamma(N_f=N_f^*)=1\label{eq:gamma-bottom}\end{equation}
where $\gamma$ is the critical exponent  
   defined as the
 derivative of the beta function at the fixed point

\begin{equation}\gamma={d\beta(\alpha)\over d\alpha}\vert_{\alpha=\alpha_{IR}}
\label{eq:exponent}\end{equation}
 As is well known, this
is a {\em universal} quantity, independant of the definition of the coupling, and  eq.(\ref{eq:gamma-bottom})
is a renormalization scheme invariant condition, as it should. The argument proceeds in two steps.

1) Let us first assume (this will be justified below) there is an 
$\alpha>\alpha_{IR}$ UV Landau singularity (for instance a pole in the beta
function at $\alpha_P>\alpha_{IR}$), in the domain of attraction of $\alpha_{IR}$.
One can then show \cite{Gar-Gru-conformal} that eq.(\ref{eq:gamma-causal})
is a 
necessary 
 condition for causality. 
I  give an improved version of the argument of \cite{Gar-Gru-conformal}. Solving the RG equation
$d\alpha/d\log k^2= \beta(\alpha)$
around $\alpha=\alpha_{IR}$, one gets
\begin{equation}\alpha(k^2)=\alpha_{IR}-\left({k^2\over\Lambda^2}\right)^{\gamma}+...
\label{eq:alpha-low}\end{equation}
There are thus rays
 \begin{equation}k^2=\vert k^2\vert\exp\left(\pm {i\pi\over\gamma}\right)\label{eq:ray}\end{equation} 
in the complex
 $k^2$ plane,
 which in the
infrared limit
$\vert k^2\vert\rightarrow 0$ are mapped by eq.(\ref{eq:alpha-low}) to positive real values of the coupling
{\em larger} then $\alpha_{IR}$. 
Assuming an expansion
\begin{equation}\beta(\alpha)=\gamma(\alpha-\alpha_{IR})+\gamma_1(\alpha-\alpha_{IR})^2+...
\label{eq:beta-low}\end{equation}
the corrections to eq.(\ref{eq:alpha-low}) are given by a series
\begin{equation}\log (k^2/\Lambda^2)={1\over \gamma} \log(\alpha_{IR}-\alpha)+{\gamma_1\over\gamma^2}
(\alpha_{IR}-\alpha)+...\label{eq:alpha-low-low}\end{equation}
with {\em real} coefficients,  showing that the only contribution to the phase for $\alpha>\alpha_{IR}$  comes from
the logarithm on the right hand side of eq.(\ref{eq:alpha-low-low}). The trajectories in the $k^2$ plane which
map  to the $\alpha>\alpha_{IR}$ region are thus straight lines to all
orders of perturbation theory around $\alpha_{IR}$. This fact suggests that
 even away from the infrared limit, 
these trajectories  are given by the rays eq.(\ref{eq:ray}). As $\vert k^2\vert$ is increased along these rays,
 the coupling will flow
  to the assumed UV
Landau singularities , reached at some finite value of $\vert k^2\vert$. If $\gamma>1$ the 
 rays, hence also the  singularities,  are located on
the first sheet of the $k^2$ plane, showing that eq.(\ref{eq:gamma-causal}) 
is a necessary condition for causality. 

2) To assert whether eq.(\ref{eq:gamma-causal}) is also sufficient  for causality, one has to make sure
 that
{\em no other sources} of (first sheet) Landau singularities
are present, but the one arising from the $\alpha>\alpha_{IR}$ region.
Barring Landau singularities at complex $\alpha$, a potential problem can still arise from an 
eventual
 UV Landau 
singularity at $\alpha<0$,
in the domain of attraction of the {\em trivial} IR fixed point $\alpha=0^-$. Indeed at weak coupling
the solution of the RG equation is controlled by the 2-loop beta function
\begin{equation}\log (k^2/\Lambda^2)={1\over \beta_0 \alpha}+{\beta_1\over \beta_0^2}\log \alpha+const+....
\label{eq:alpha-2loop}\end{equation}
where the $const$ is real. For $\alpha<0$ the right hand side of eq.(\ref{eq:alpha-2loop}) acquires a 
$\pm i\pi{\beta_1\over \beta_0^2}$ imaginary part, which implies   the rays
\begin{equation}k^2=\vert k^2\vert\exp\left(\pm i\pi{\beta_1\over \beta_0^2}\right)\label{eq:alpha<0}\end{equation}
map to the $\alpha<0$ region.
Along the rays eq.(\ref{eq:alpha<0}), we are effectively in a QED like situation:
increasing $\vert k^2\vert$, the coupling is either attracted to a non-trivial UV fixed point, or reaches an UV
Landau singularity at some finite $\vert k^2\vert$. In the latter case,
one must  require that
the condition:
\begin{equation}\vert{\beta_0^2\over \beta_1}\vert<1\label{eq:neg-landau-causal}
\end{equation}
 is satisfied in the whole $N_f$ range where eq.(\ref{eq:gamma-causal}) is valid, which 
will  confine  the rays, hence the singularities to the second (or higher) sheet.

However in QCD condition eq.(\ref{eq:neg-landau-causal}) can be satisfied
{\em only} if $\beta_1<0$,
and   coincides\footnote{Eq.(\ref{eq:neg-landau-causal}) is however
 a necessary condition for causality for {\em any} beta function which admits an  UV Landau singularity at negative $\alpha$
 (in the domain
of attraction of the $0^-$ trivial IR fixed point),
 and applies also if $\beta_1$
is positive in a general theory!} with the 2-loop causality condition 
eq.(\ref{eq:gamma-2loop}), 
  which requires $N_f>9.7$. Therefore, to preserve causality within
 the conformal window as determined\footnote{It is a priori possible in QCD to have an $\alpha<0$ UV Landau singularity 
rather then an $\alpha<0$ UV fixed point. In such a case
the bottom of the conformal window would be given as in the two-loop model by the condition $-\beta_0^2/\beta_1=1$, yielding
$N_f^*\simeq 9.7$. Indeed, at such large $N_f$, any eventual UV Landau singularity from the $\alpha>\alpha_{IR}$ region has not
yet reached the first sheet since $\gamma<0.4$  as shown by Fig. 1. This possibility is however disfavored as explained in the text.} by eq.(\ref{eq:gamma-causal}),  $\gamma$ should reach $1$ in the region $N_f>9.7$,
 which is clearly excluded
 (see Fig. 1 below). In order that eq.(\ref{eq:gamma-causal}) be also a sufficient condition for causality
one must thus check   that a {\em non-trivial}
(finite or infinite) UV 
fixed point $\alpha_{UV}$ is present 
at {\em negative} $\alpha$!
A minimal example satisfying this requirement is the  3-loop beta function 
$\beta(\alpha)=-\beta_0 \alpha^2-\beta_1 \alpha^3-\beta_2 \alpha^4$
with $\beta_0>0$ and $\beta_2<0$ ($\beta_1$ can have any sign).

It is  worth mentionning  eq.(\ref{eq:neg-landau-causal}) is
{\em always}
violated \cite{Gar-Gru-conformal} in the lower part of the conformal window in  SQCD as determined by duality \cite{Seiberg},
 and the 
previous argument
 thus implies the existence
 of a negative UV fixed point in this theory. In fact the ``exact'' NSVZ \cite{NSVZ} beta function for $N_f=0$
 does exhibit  an (infinite) UV fixed point as $\alpha\rightarrow -\infty$, which might
be the parent of a similar one present within Seiberg conformal window.

There is some evidence that a negative UV fixed point is indeed also present in QCD. At the three-loop level, QCD effective
charges associated to Euclidean correlators 
have typically $\beta_2<0$, and
 appear  \cite{Gar-Gru-conformal} to be causal and  admit a negative
UV fixed point,  even somewhat {\em below} the two-loop
causality boundary $N_f = 9.7$. This evidence could be washed out in yet higher orders (e.g. the Pade improved
three-loop beta functions \cite{Gar-Gru-conformal}). However, a different and more sytematic derivation is given
 in section 7. The plausible existence  of a nearby negative UV fixed point\footnote{The non-trivial UV fixed point is actually not relevant to the proper analytic continuation of the 
coupling at complex $k^2$, which must be consistent \cite{Gar-Gru-Kar} with (UV) asymptotic freedom. This means that in presence
of this fixed point, the correct analytic continuation must involve {\em complex} rather then negative
values of $\alpha$ along the rays eq.(\ref{eq:alpha<0}), and one should  approach
 the non-trivial
(rather then the trivial) IR fixed point as $\vert k^2\vert\rightarrow 0$, 
and the trivial (rather then the non-trivial) UV fixed point as $\vert k^2\vert\rightarrow \infty$. This is possible since the
solution of eq.(\ref{eq:alpha-2loop}) is not unique for a given (complex) $k^2$.
 For the same
 reason, any eventual IR
Landau singularity arising from the region $\alpha<\alpha_{UV}$, in the domain of attraction
of the non-trivial UV fixed point, is  not relevant to the correct analytic continuation. On the other hand, any UV Landau
singularity in the domain of attraction of either the trivial or the non-trivial IR fixed points as considered
above {\em is relevant} to the
proper analytic continuation, since the coupling will flow to the only UV fixed point available, namely the {\em trivial}
one, once the UV Landau singularity is passed.
} in QCD, hence the absence  of
 an UV  Landau singularity at negative $\alpha$, justifies
 a posteriori (barring complex $\alpha$ singularities)
the above assumption that there must be an $\alpha>\alpha_{IR}$ singularity, to provide the necessary  causality violation 
below the conformal window.

\section{Computing $N_f^*$ through the Banks-Zaks expansion}
 One can  try to use
 the Banks-Zaks
expansion to compute $\gamma$ and determine $N_f^*$, the lower boundary of the conformal window.
The Banks-Zaks expansion \cite{BZ,White,G-bz} is an expansion of the fixed point in powers of the distance $N_f^0-N_f=16.5-N_f$
from the top of the conformal window, which is  proportional to $\beta_0$. The solution of the equation

\begin{equation}\beta(\alpha)=-\beta_0 \alpha^2-\beta_1 \alpha^3-\beta_2 \alpha^4+...=0\label{eq:fixed-point}
\end{equation}
in the limit  $\beta_0\rightarrow 0$, with $\beta_i$ ($i\geq 1$) finite is obtained as a power series

\begin{equation}\alpha_{IR}=a+{\cal O}(a^2)\label{eq:BZ}
\end{equation}
where the expansion parameter $a\equiv {8\over 321}(16.5-N_f)={16\over 107}\beta_0$.
The Banks-Zaks expansion for the critical exponent eq.(\ref{eq:exponent}) is presently
 known \cite{Stev,Gar-Kar,Gar-Gru-conformal} up 
to next-to-next-to leading order :

\begin{equation}\gamma={107\over 16}a^2(1+4.75 a-8.89 a^2+...)
\label{eq:exponent-BZ}\end{equation}
Using the trunkated expansion eq.(\ref{eq:exponent-BZ}), one finds that $\gamma<1$ for $N_f\geq 5$,
with $\gamma=1$ reached for  $N_f=N_f^*\simeq 4$. To assess whether it is reasonable to use 
perturbation theory down to $N_f=N_f^*$, let us look at the magnitude of the successive terms within
the parenthesis in eq.(\ref{eq:exponent-BZ}). They are given by:
$1, 1.44, -0.82$. Although the next to leading term gives a very large correction, and the series
seem at best poorly converging at $N_f=N_f^*$, one can observe that the 
next-to next to leading term still gives a moderate correction to the sum of the first two terms, which
might be considered together \cite{Gar-Gru-conformal} as building the ``leading'' contribution,
 since they are
 both derived from information
contained \cite{G-bz,Gar-Gru-conformal} in the minimal 2-loop beta function necessary to get a non-trivial fixed point.
 Indeed, keeping only the first two terms
in eq.(\ref{eq:exponent-BZ}), one finds that $\gamma=1$ is reached for  $N_f=N_f^*\simeq 6$. On the
other hand, using a $[1,1]$ Pade approximant as a model\footnote{The alternative $[0,2]$ Pade yields a result ($\gamma<0.26$ for any $N_f$)
 inconsistent with the present framework. It also predicts a not very plausible ${\cal O}(a^5)$
coefficient of $\simeq -192$.}
 for extrapolation of the perturbative
series, one gets

\begin{equation}\gamma={107\over 16}\ a^2\ {1+6.62 a\over 1+1.87 a}\label{eq:exponent-Pade}
\end{equation} 
which yields  $\gamma=1$ for $N_f=N_f^*\simeq 5$. Fig. 1 shows  $\gamma$ as a function
of $N_f$: \EPSFIGURE{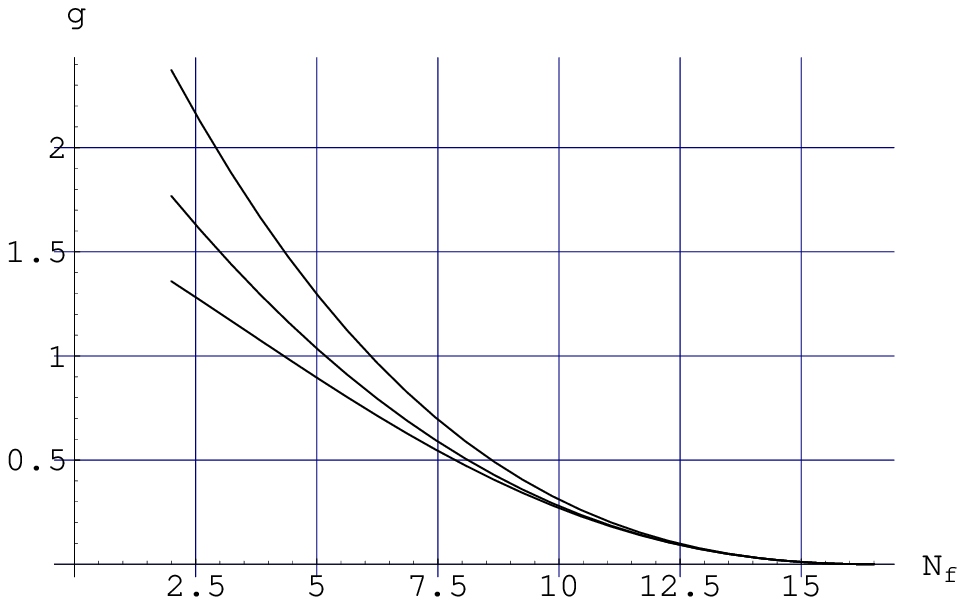}{The critical exponent as a function of $N_f$:
 top: ${\cal O}(a^3)$ order; middle: Pade; bottom: ${\cal O}(a^4)$ order.}

%\begin{figure}[H]
%\begin{center}
%\mbox{\kern-0.5cm
%\epsfig{file=gamma1.ps,width=10.0truecm,angle=0}}
%\end{center}
%\caption{The critical exponent as a function of $N_f$:
% top: ${\cal O}(a^3)$ order; middle: Pade; bottom: ${\cal O}(a^4)$ order.}
%\label{pert}
%\end{figure}

%\epsfbox{gamma1.ps}

%\newpage

Note that in the obtained range of $N_f^*$ values
($4<N_f^*<6$),
$\beta_1$ is still positive  ($\beta_1$ changes sign for $N_f = 8.05$)
and of the same sign as $\beta_0$. The fixed point must thus arise
from the contributions of higher then 2 loop beta function corrections,
 although I am assuming the Banks-Zaks 
expansion is still converging there. This is consistent with the previously mentionned fact that QCD effective
charges have
{\em negative} 3-loop beta function coefficients in the above range of $N_f$ values.

\section{Further evidence for a negative UV fixed point in QCD}
 Additional evidence for the existence
of a {\em couple} of (negative-positive) UV-IR fixed points  is provided by the following modified Banks-Zaks argument.
Assume $\beta_1=0$, i.e. $N_f=8.05$.  Then a real
 fixed point can
 still exist at the three-loop level
if $\beta_2<0$, and actually one gets a {\em pair} of opposite signs zeroes at ${\tilde \alpha}= \pm (-\beta_0/\beta_2)^{1/2}$,
 an IR
and an UV fixed point. If $\beta_0$ is small enough, they are weakly coupled, and calculable through a modified
Banks-Zaks expansion around $N_f=16.5$,  applied to the auxiliary function
 ${\tilde \beta}(\alpha)\equiv  \beta(\alpha)+\beta_1\alpha^3$ with the two-loop term removed. One gets

\begin{equation}{\tilde \alpha}=\pm {\tilde a}\left(1 \mp {1\over 2}\ {\beta_{3,0}\over \beta_{2,0}}\ {\tilde a}+...\right)
\label{eq:BZ2}\end{equation} 
where the expansion parameter ${\tilde a}\equiv (-\beta_0/\beta_{2,0})^{1/2}$, and $\beta_{i,0}$,  the $N_f=16.5$ values
of $\beta_i$ ($i=2,3$), are scheme dependent and can be obtained \cite{G-bz} from the knowledge of the coefficients in eq.(\ref{eq:exponent-BZ}).
 For  effective charges associated to Euclidean correlators,
 the correction in eq.(\ref{eq:BZ2}) ranges 
from $0.1$ to $0.7$ at $N_f=8.05$ (where ${\tilde \beta}(\alpha)$ coincides with  $ \beta(\alpha)$),
 which is encouraging
evidence for a couple of UV and IR fixed points around this value of $N_f$ (which is within the alleged
conformal window, but below the 2-loop causality region).  Additional support is given by the calculation 
of the auxiliary critical exponent
${\tilde \gamma}\equiv
d{\tilde\beta}(\alpha)/ d\alpha\vert_{\alpha={\tilde\alpha_{IR}}}$, whose modified Banks-Zaks expansion is

\begin{equation}{\tilde \gamma}= 2\ \beta_0\ {\tilde a}(1+{\cal O}({\tilde a}^2))\label{eq:BZ2-exponent}\end{equation}
(no ${\cal O}({\tilde a})$ correction!), which yields 
 at $N_f=8.05$ (where it coincides with  $\gamma$)
$0.6<{\tilde \gamma}<0.7$ for  effective charges associated to Euclidean correlators, in reasonable agreement with the standard
 Banks-Zaks result (Fig. 1) $0.5<\gamma<0.6$.

\section{Conclusions}

1) I have suggested a direct, {\em physical} motivation for Landau singularities, assuming a two-phase
structure of QCD: they should trace the lower boundary $N_f^*$ of the conformal window. This approach avoids
the notoriously tricky disentangling of the ``perturbative'' from the ``non-perturbative'' part of the QCD
amplitudes within the conformal window, since they are by definition entirely ``perturbative'' there. On the other
hand, such a disentangling is naturally achieved below the conformal window, by introducing the analytic continuation
of the conformal window amplitudes to the $N_f<N_f^*$ region.

2) Assuming that the {\em perturbative} QCD coupling has a non-trivial IR fixed point $\alpha_{IR}$
 even in some range {\em below} $N_f^*$ leads
to the equation $\gamma(N_f=N_f^*)=1$ to determine $N_f^*$ from the critical exponent $\gamma$ at the IR fixed point.
Using the available terms in the Banks-Zaks expansion, this equation yields $4\leq N_f^*\leq 6$. It would clearly be desirable to have more terms to better control the
accuracy of the Banks-Zaks expansion.  Note that this condition
is inconsistent with the existence of an analogue of Seiberg duality  in QCD,
 which would rather imply  $\gamma(N_f=N_f^*)=0$.

3) Some conditions on the QCD beta function are required: the {\em only} source of (UV) Landau singularities must
come from the $\alpha>\alpha_{IR}$ region. One needs in
particular
a {\em negative} UV fixed point $\alpha_{UV}$. There is indeed some evidence for such a fixed point in QCD.

4) A  negative UV fixed point is also required in the SQCD case, 
where duality fixes  the conformal window.
 
5) It is possible the {\em finite} IR fixed point  persists in the perturbative QCD coupling down to the $N_f\rightarrow -\infty$ one-loop limit. A simple
example is provided by a beta function with one positive pole $\alpha_P$ (the required Landau singularity) and two 
opposite sign
 zeroes $\alpha_{IR}$ and $\alpha_{UV}$:
$\beta(\alpha)=-\beta_0\alpha^2(1-\alpha/ \alpha_{IR}) (1-\alpha/ \alpha_{UV})/
 (1-\alpha/ \alpha_P)$
where $\alpha_{UV}<0$ and $0<\alpha_{IR}<\alpha_P$. The one-loop limit is achieved for $\alpha_{IR}=\alpha_P$
and $\alpha_{UV}=-\infty$.
A paper developing further these issues is under preparation.

\acknowledgments
I thank A. Armoni and A.H. Mueller for  discussions.

%\newpage

% Upper-case    A B C D E F G H I J K L M N O P Q R S T U V W X Y Z
% Lower-case    a b c d e f g h i j k l m n o p q r s t u v w x y z
% Digits        0 1 2 3 4 5 6 7 8 9
% Exclamation   !           Double quote "          Hash (number) #
% Dollar        $           Percent      %          Ampersand     &
% Acute accent  '           Left paren   (          Right paren   )
% Asterisk      *           Plus         +          Comma         ,
% Minus         -           Point        .          Solidus       /
% Colon         :           Semicolon    ;          Less than     <
% Equals        =           Greater than >          Question mark ?
% At            @           Left bracket [          Backslash     \
% Right bracket ]           Circumflex   ^          Underscore    _
% Grave accent  `           Left brace   {          Vertical bar  |
% Right brace   }           Tilde        ~

\end{document}